\documentclass[12pt,preprint]{aastex}

\newcommand{\etal}{et~al.\ }
\newcommand{\kms}{\hbox{km~s$^{-1}$}}
\newcommand{\cmsq}{\hbox{cm$^{-2}$}}

\newcommand{\flux}{\hbox{erg~cm$^{-2}$~s$^{-1}$}}
\newcommand{\lumin}{\hbox{erg~s$^{-1}$}}

\newcommand{\nh}{\hbox{${N}_{\rm H}$}}

\newcommand{\chandra}{{\emph{Chandra}}}

\newcommand{\asca}{{\emph{ASCA}}}

\newcommand{\hst}{\emph{HST}}

\newcommand{\bzero}{B1600+434}
\newcommand{\beight}{B1608+656}

\begin{document}

\def\sarc{$^{\prime\prime}\!\!.$}
\def\arcsec{$^{\prime\prime}$}
\def\arcmin{$^{\prime}$}
\def\degr{$^{\circ}$}
\def\seco{$^{\rm s}\!\!.$}
\def\ls{\lower 2pt \hbox{$\;\scriptscriptstyle \buildrel<\over\sim\;$}} 
\def\gs{\lower 2pt \hbox{$\;\scriptscriptstyle \buildrel>\over\sim\;$}} 
 
\title{\chandra\ Observations of Gravitational Lenses B1600+434 and B1608+656}

\author{Xinyu Dai\altaffilmark{1} and Christopher S. Kochanek\altaffilmark{1}}

\altaffiltext{1}{Department of Astronomy,
The Ohio State University, Columbus, OH 43210,
xinyu@astronomy.ohio-state.edu, ckochanek@astronomy.ohio-state.edu}

\begin{abstract}
We observed \bzero\ and \beight\ with CXO/ACIS, detecting both quasar images in \bzero\ and three of four images in \beight.
We did not detect significant X-ray emission from nearby galaxy groups or clusters associated with each lens galaxy. The upper limits on the X-ray luminosity of any cluster within 4\arcmin\ of each lens and at each lens redshift are of $\sim2\times10^{42}$ and $\sim6\times10^{42}$~\lumin\ for \bzero\ and \beight, respectively.  
The radio-loud source quasars have power-law photon indices of $\Gamma=1.9\pm0.2$ and $\Gamma=1.4\pm0.3$ and X-ray luminosities of $1.4^{+0.2}_{-0.1}\times$10$^{45}$ and $2.9^{+0.7}_{-0.4}\times10^{44}$~\lumin\ for \bzero\ and \beight, respectively before correcting for the magnification.  
We detected a differential absorption column density of $\Delta\nh\sim3\times10^{21}\cmsq$ between the two images of \bzero, roughly consistent with expectations from differential extinction estimates of $\Delta E(B-V)=0.1$~mag and a standard dust-to-gas ratio.  
The differential absorption observed in gravitational lenses may serve as an important probe to study the gas content in high redshift galaxies since it can separate the absorbing column originating from the lens galaxy and those intrinsic to quasars.  
We also detected 157 serendipitous X-ray sources in the two \chandra\ fields and identified the brighter optical counterparts using the SDSS and DPOSS surveys.
\end{abstract}


\section{Introduction}
The radio-loud gravitational lenses \bzero\ and \beight\ were discovered in the Cosmic Lens All Sky Survey \citep{ja95,my95}.  \beight\ was also discovered independently by \citet{sn95}.
\bzero\ is a two image system separated by 1\sarc4 with source redshift $z=1.59$ \citep{ja95,fc98}.  
The lens galaxy of \bzero\ is an edge-on early-type spiral galaxy at $z=0.41$ \citep{jh97,fc98}.
\beight\ is a four image system about 2\sarc1 across with a source redshift at $z=1.39$ \citep{fa96} and a lens redshift at $z=0.63$ \citep{my95}.
The lens of \beight\ consists of two interacting elliptical galaxies within the Einstein ring.
In addition, \hst\ observations have detected extended images of the AGN host galaxies in both systems \citep{ja98,koch99,kkm01,sb03}.

One of the important applications of gravitational lenses is to constrain the Hubble constant \citep{r64} or alternatively, the dark matter distribution in the lens galaxy, through time-delay measurements between the lensed images.  
The time-delays in both \bzero\ and \beight\ have been measured \citep{fa99,fa02,ko00,bu00}.
Modeling of the two gravitational lenses has been carried out by several studies (e.g., Koopmans \& Fassnacht 1999; Koopmans \etal 2003; Kochanek 2002, 2003).
In most studies, there are model degeneracy problems in that a number of models can reproduce the limited number of constraints and predict different Hubble constant values based on each time delay measurement.
Therefore, it is important to map the mass distribution close to the lens galaxy with other independent methods to better constrain the lens models.
In many lenses, galaxy groups or clusters close to the lenses provide non-negligible contribution to the lens potential \citep{kz04,fl02}.
X-ray observations are particularly important in this respect because they can accurately constrain the positions and masses of the galaxy groups or clusters.
In particular,
the centroid of the X-ray emission from the intracluster gas provides a more accurate measurement of the cluster's position than optical studies.
X-ray clusters or groups have been observed in the lenses RX~J0911.4+0551, Q0957+561, B1422+231, and PG~1115+080 \citep{mo01,ch02c,rsw03,gr04}.

The superb angular resolution of \chandra\ also allows us to resolve the lensed quasar images if they are separated by more than 0\sarc5.
Thus, besides constraining the properties of galaxy groups or clusters, X-ray observations of gravitational lenses can also be used to constrain the sizes of quasar X-ray emission regions \citep{da03,ch02a,ch04}, to measure ultra-short time delays between lensed images \citep{ch01,da03,cdg04}, and to study the interstellar medium of the lens galaxies (e.g., Dai \etal 2003).
In addition, the flux magnification provided by gravitational lensing facilitates studies of objects that are intrinsically X-ray faint, such as broad absorption line quasars \citep{ch02b,ga02} and high redshift quasars \citep{da04}.

In this paper, we present results from \chandra\ observations of \bzero\ and \beight.
We discuss the data acquisition and processing in $\S$2, the spectral properties of the lensed quasars in $\S$3, our limits on the presence of X-ray luminous groups or clusters in $\S$4, and the serendipitous sources in the fields in $\S$5.
We assume cosmology with $H_0 = 70~\rm{km~s^{-1}~Mpc^{-1}}$, $\Omega_{\rm m} = 0.3$, and $\Omega_{\Lambda}= 0.7$ throughout the paper.

\section{Observations and Data Reduction}
We observed \bzero\ and \beight\ with the Advanced CCD Imaging Spectrometer (ACIS; Garmire \etal 2003) onboard \chandra\ \citep{we02} for $\sim30.2$~ks and $\sim29.7$~ks on 2003 October 7, and 2003 September 21, respectively.  
The data were taken continuously with no interruptions during each observation, and there were no significant background flares during the observations.
Both gravitational lenses were placed on the back-illuminated ACIS-S3 chip, where the aim points of the observations are located, and the data were taken in the TIMED/VFAINT mode.
\bzero\ was placed near the aim point of the observation ($\sim$8\arcsec\ off-axis) and \beight\ was placed much further from the aim point about $\sim$75\arcsec\ off-axis.
The latter lens was shifted off-axis to ensure a galaxy clump in the \hst\ images would stay in the field.
This configuration was used to maximize the sensitivity for detecting nearby galaxy groups or clusters that might contribute to the lensing and affect estimates of the Hubble constant.
The data were reduced with the \verb+CIAO 3.1+ software tools provided by the \emph{Chandra X-ray Center} (CXC).
We improved the image quality of the data by removing the pixel
randomization applied to the event positions
in the CXC processing and by applying a subpixel resolution
technique \citep{t01,m01}.
In the data analysis, only events with standard \asca\ grades of 0, 2, 3, 4, and 6 were used.

\section{Lensed Quasars}
We first discuss the X-ray properties of the lensed quasars in both lens systems.
The imaging and spectral analysis are discussed in $\S$\ref{sec:img} and $\S$\ref{sec:spec} and the X-ray flux ratios of the lensed images are estimated in $\S$\ref{sec:fr}.
\subsection{Imaging Analysis\label{sec:img}}
\begin{figure}
\epsscale{1}
\plotone{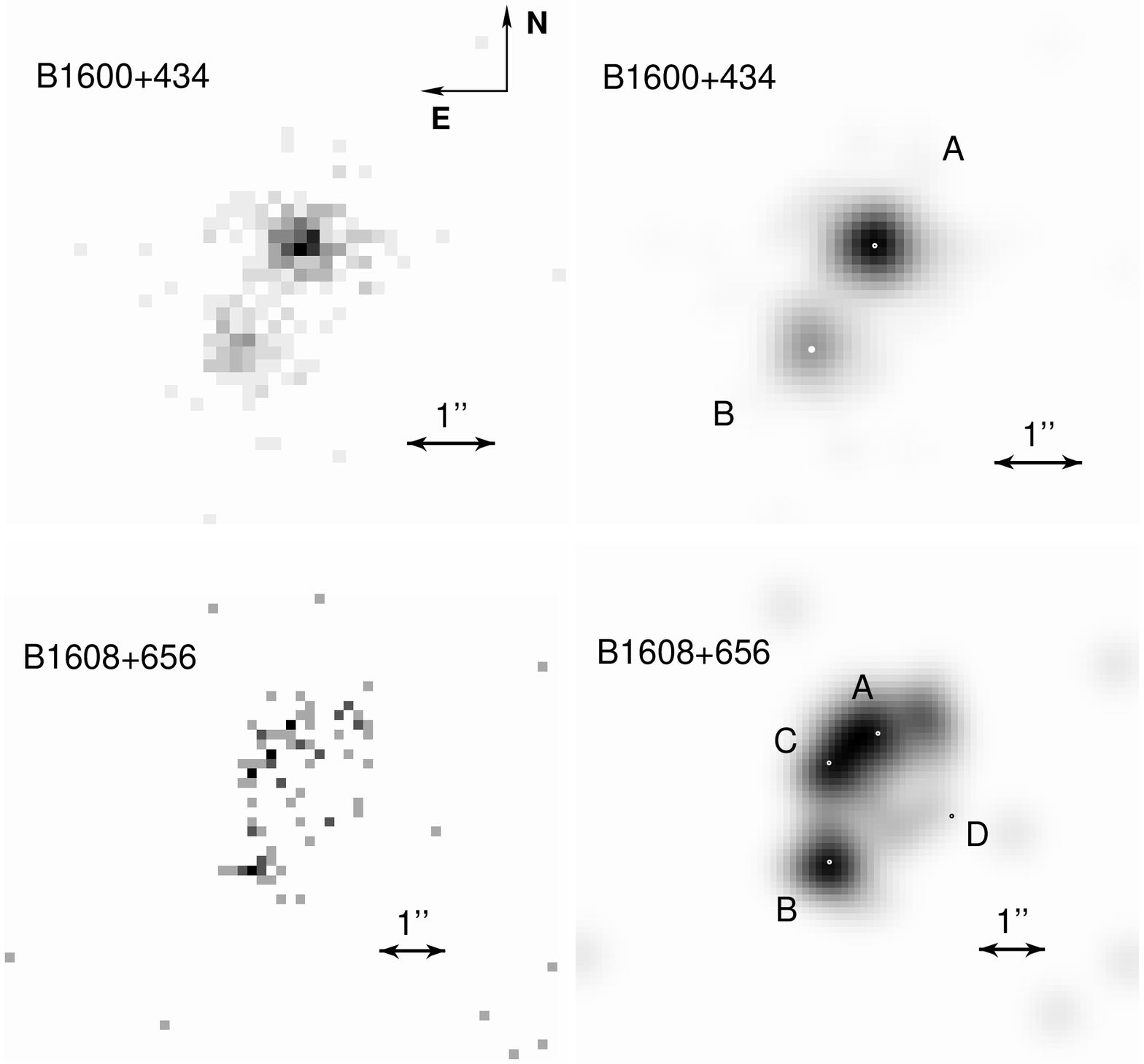}
\caption{In the left panel we show the raw images of \bzero\ (top) and \beight\ (bottom) with the events binned in 0\sarc15 bins.  In the right panel we have smoothed the images with an 0\sarc2 Gaussian for \bzero\ (top) and an 0\sarc3 Gaussian for \beight\ (bottom).  The small circles in the smoothed images show the image positions measured from the CASTLES \hst\ images of the systems.  The scales of the smoothed and unsmoothed images are the same.  We note that the 75\arcsec\ off-axis angle of \beight\ observation does not affect the image analysis as the 50\% encircled energy contour is of 0\sarc5 radius as opposed to 0\sarc4 on-axis.
\label{fig:img}}
\end{figure}

We analyzed images limited to the 0.2--8 keV band where we detected a total
of 332 and 84 net events for \bzero\ and \beight, respectively.
The raw and smoothed \chandra\ images of \bzero\ and \beight\ are shown in Figure~\ref{fig:img}.  
Both of the raw images are binned with a bin size of 0\sarc15.
The smoothed images are generated by binning the images with a bin size of 0\sarc1 for both lenses and then were smoothed with Gaussians of width $\sigma=$~0\sarc2 and $\sigma=$~0\sarc3 for \bzero\ and \beight, respectively.

In \bzero\ we detected both lensed images.
The measured image separation (1\sarc35$\pm$0\sarc05) is consistent with the separation measured in the CfA-Arizona Space Telescope LEns Survey (CASTLES) \hst\ observations.
\footnote{The CASTLES website is located at http://cfa-www.harvard.edu/castles/.}

The situation is more complicated in \beight\ due to the large number of images with smaller minimum separations, but also because the count rate of the lens is lower.
The relatively larger off-axis angle does not degrade the point spread function (PSF) very much.
We simulated the PSF using the \verb+ChaRT+ \citep{ca03} and \verb+MARX+ \citep{wi97} tools and found the 50\% encircled energy contour is of $\sim$0\sarc5 radius as opposed to $\sim$0\sarc4 on axis.
Three out of four lensed images are clearly detected.
Image B is well separated, but images A and C are merged.
We fit the image assuming there are three point images of A, B, and C constraining the relative quasar position to match the CASTLES $H$-band relative position and using the above mentioned PSF model.
However, we are unable to obtain a stable solution.
This may be due to the low count-rate of the source or it may indicate that the images are not point sources.
We note that there are some X-ray events about 1\arcsec\ Northwest to image A and we are unable to determine whether these events belong to image A or from other unidentified sources.
The \hst\ images do not show any source at the corresponding position other than part of the Einstein ring image of the host galaxy.
Image D is not detected by the \chandra\ observation.
The non-detection of image D could be caused by small number statistics -- additional absorption is not required to mask the image. 

\subsection{Spectral Analysis\label{sec:spec}}
We extracted spectra for each quasar using the \verb+CIAO+ tool
\verb+psextract+ and fit them using \verb+XSPEC V11.3.1+ \citep{a96}
within the 0.4--8 keV observed energy range.
We note that in \verb+CIAO 3.1+ the recently observed decline in the low energy quantum efficiency of ACIS, possibly caused by molecular contamination of the ACIS filters, is already accounted for by the \verb+psextract+ tool.  Therefore, we did not perform any other corrections beyond that.
\subsubsection{\bzero}
We extracted both the combined spectrum of both images within a circle with 5\arcsec\ radius about the centroid of the combined image and individual spectra of images within 0\sarc7 radius circles about the centroids of images A and B, respectively.
We extracted the background spectrum within an annulus with inner and outer radii of 10\arcsec\ and 30\arcsec.
We fitted the spectra of \bzero\ with a series of models summarized in Table~\ref{tab:spec}.
We first modeled the spectrum as a power-law modified by Galactic absorption.
This model gives a reasonable fit with $\chi^2=16.3$ for 18 degrees of freedom (dof).
However, the column density obtained for Galactic absorption, \nh~$=(14\pm4)\times10^{20}$~cm$^{-2}$, is much higher than the expected \nh~$=1.3\times10^{20}$~cm$^{-2}$ \citep{d90}. 
This exercise indicates that additional absorption components are needed to interpret the spectrum.  
These additional absorption components can be either intrinsic to the quasar, located in the quasar host galaxy, or in the lens galaxy.  
We found that adding a neutral absorption component at the redshift of either the quasar or the lens can produce an acceptable fit
with $\chi^2=16.1$ for 18 dof and $\chi^2=15.8$ for 18 dof, respectively, for absorption at the quasar and lens redshifts.
The absorption column densities obtained are \nh~$=(90\pm40)\times10^{20}$~cm$^{-2}$ and \nh~$=(22\pm8)\times10^{20}$~cm$^{-2}$ for absorption at the quasar or lens redshift, respectively.
The combined spectrum and the best fit model with absorption at the lens redshift are shown in Figure~\ref{fig:spec}a.
The estimate of the intrinsic photon index is around $\Gamma\sim1.9$ for both models with observed 0.4--8 keV band flux of $6.9^{+0.7}_{-1.1}\times$10$^{-14}$~\flux, and lensed, unabsorbed X-ray luminosities in the 2--10 keV and 1--20 keV band rest frame of $7.6^{+1.6}_{-0.2}\times$10$^{44}$ and $1.4^{+0.2}_{-0.1}\times$10$^{45}$~\lumin, respectively.
We note that the lensing magnification must be included in order to obtain the unlensed luminosity.     

\begin{figure}
\epsscale{1}
\plotone{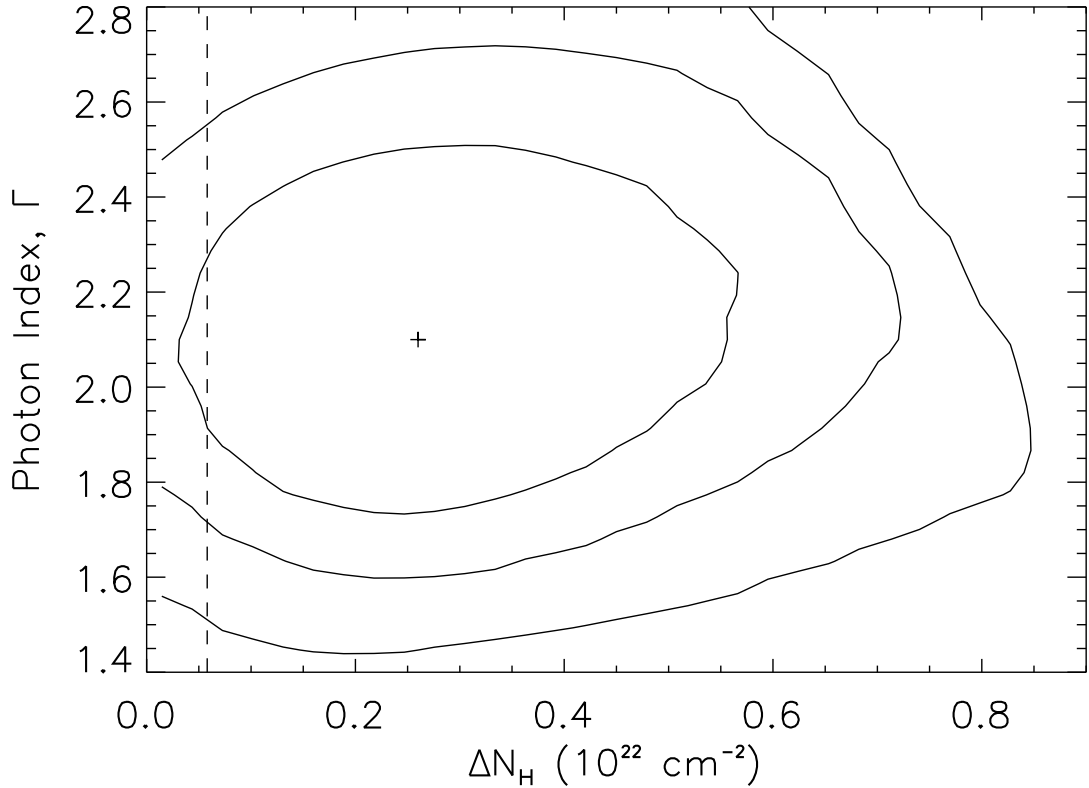}
\caption{68\%, 90\% and 99\% confidence contours on the photon index and the differential absorption column density $\Delta$\nh\ at the lens ($z=0.41$) between images A and B of \bzero. The differential absorption is detected at approximate 90\% confidence.  The dashed line indicates the differential absorption expected from the differential extinction of $\Delta E(B-V)=0.1$~mag between images A and B \citep{fal99} and assuming a standard dust-to-gas ratio of $5.8\times10^{21}~{\rm mag^{-1}~cm^{-2}}$ (Bohlin \etal 1978). \label{fig:con}}
\end{figure}

\citet{fal99} interpreted the wavelength dependence of the image flux ratios as due to $\Delta E(B-V)=0.1$ of differential extinction obscuring image B.
We also see in the spectra of the image that image B has proportionally less soft X-ray emission so that differential absorption must be intrinsic to the lens.
To estimate this differential absorption we simultaneously fit
the individual spectra of images A and B constrained to have the
same intrinsic power-law photon index and Galactic absorption column density but allowed to have different absorption column densities at the redshift of the lens.  
The best fit model is shown in Figure~\ref{fig:spec}b.  
This model also results in acceptable fit with a $\chi^2=13.1$ for 12 dof as compared to $\chi^2=16.4$ for 13 dof if we do not allow for differential absorption.
The improvement of the fit by allowing for differential absorption is significant at the 90\% confidence level according to the F-test
\footnote{We did not probe the boundary of the parameter space, where the F-test fails to apply \citep{pr02}.} indicating that the X-ray data are consistent with, but do not require, column density variations.
The photon index obtained in this fit is slightly steeper than for the previous fits in part because the
small extraction regions used to obtain individual spectra of the images to the loss of some hard X-ray photons.
The absorption column densities at the lens for images A and B are \nh~$=(21\pm11)\times10^{20}$~cm$^{-2}$ and \nh~$=47^{+21}_{-17}\times10^{20}$~cm$^{-2}$, respectively, and the differential absorption between the two images is $\Delta\nh\sim3\times10^{21}\cmsq$ as shown in Figure~\ref{fig:con}.
This indicates that the \nh\ determined from the two spectra are different at the 90\% confidence level with higher \nh\ for image B.
This is consistent with the lens configuration, where the lens galaxy is much closer to image B than to image A.
Comparing with the estimated differential extinction of $\Delta E(B-V)=0.1$~mag between images A and B \citep{fal99}, we obtained a dust-to-gas ratio of $2.6^{+1.9}_{-1.5}\times10^{22}~{\rm mag^{-1}~cm^{-2}}$ within a factor of five from the Galactic value, $5.8\times10^{21}~{\rm mag^{-1}~cm^{-2}}$ \citep{bsd78}.

\subsubsection{\beight} 
\begin{figure}
\epsscale{0.6}
\plotone{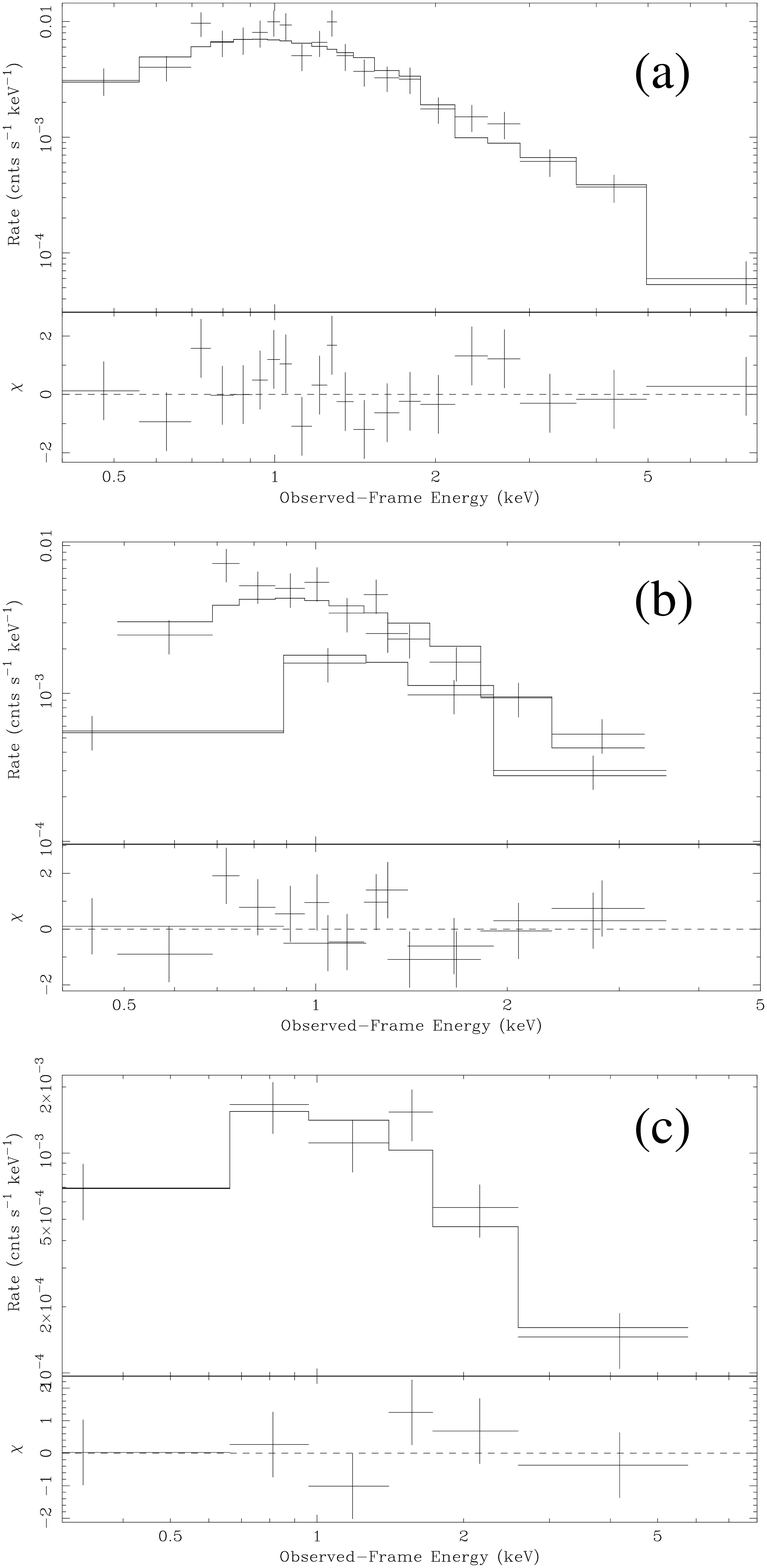}
\caption{\chandra\ spectral fitting results for \bzero\ and \beight.  (a) The combined spectrum of images A and B of \bzero. (b) Simultaneous fits to the individual spectra of images A and B of \bzero. (c) The combined spectrum of all images of \beight.\label{fig:spec}}
\end{figure}

We extracted the combined spectrum of all the images of \beight\ within a 5\arcsec\ radius circle centered on the source.
We extracted the background spectrum within an annulus with inner and outer radii of 10\arcsec\ and 18\arcsec.
The background subtracted spectrum consists of only 86 net events with a count-rate of 2.9$\times$10$^{-3}$~count~s$^{-1}$, making a spectral analysis of the individual images impossible.
We modeled the spectrum with a power-law modified by Galactic absorption.
This simple model fits the spectrum quite well, with $\chi^2=3.3$ for 3 dof.  The parameters are presented in Table~\ref{tab:spec} and the spectrum and the best fit model are shown in Figure~\ref{fig:spec}c.
The power-law photon index and Galactic \nh\ are constrained to be $\Gamma=1.4\pm0.3$ and \nh~$<6.4\times10^{20}$~cm$^{-2}$
consistent with the expected Galactic column density of \nh~$=2.7\times10^{20}$~cm$^{-2}$ \citep{d90}.
If we fix the Galactic absorption to this value the fits are consistent with no additional absorption at either the lens or quasar redshift with upper limits on absorption column of \nh~$<10^{21}$ and $<2\times10^{21}$~cm$^{-2}$ at the redshift of lens or quasar, respectively.
There is differential extinction in \beight\ as well, but it is most significant for image D which we failed to detect.
Our upper limit on the flux of D is not strong enough to estimate a dust-to-gas ratio.
The X-ray flux in the observed 0.4--8 keV band is $2.3\pm0.5\times$10$^{-14}$~\flux, and the lensed, unabsorbed X-ray luminosities in the 2-10 keV and 1--19 keV band rest frame are $1.5^{+0.4}_{-0.2}\times$10$^{44}$ and $2.9^{+0.7}_{-0.4}\times$10$^{44}$~\lumin, respectively.

\begin{deluxetable}{cccccccc}
\tabletypesize{\scriptsize}
\tablecolumns{8}
\tablewidth{0pt}
\tablecaption{Results of Fits to the \chandra\ Spectra
of \bzero\ and \beight\label{tab:spec}}
\tablehead{
\colhead{} &
\colhead{} &
\colhead{} &
\colhead{} &
\colhead{Galactic \nh} &
\colhead{Other Absorption \nh\tablenotemark{a}} &
\colhead{Flux\tablenotemark{b}} &
\colhead{}
\\
\colhead{Fit} &
\colhead{Quasar} &
\colhead{Image} &
\colhead{$\Gamma$} &
\colhead{($10^{20}~\rm{cm^{-2}}$)} &
\colhead{($10^{20}~\rm{cm^{-2}}$)} &
\colhead{($10^{-14}~\rm{erg~s^{-1}~cm^{-2}}$)} &
\colhead{$\chi^{2}(\nu)$} 
}

\startdata
1 & \bzero & Total & $2.0\pm0.2$ & $14\pm4$ & \nodata & $6.8\pm0.9$ & 16.3(18) \\
2 & \bzero & Total & $1.9\pm0.2$ & 1.29 (fixed) & $90\pm40$ $(z=1.59)$ & $6.8\pm1.0$ & 16.1(18) \\
3 & \bzero & Total & $1.9\pm0.2$ & 1.29 (fixed) & $22\pm8$ $(z=0.41)$ & $6.8\pm1.2$ & 15.8(18) \\
4\tablenotemark{c} & \bzero & A & $2.1\pm0.2$ & 1.29 (fixed) & $21\pm11$ $(z=0.41)$ & $3.6\pm0.7$ & 13.1(12) \\
  &        & B &             &      & $47^{+21}_{-17}$ $(z=0.41)$ & $1.8\pm0.6$ &      \\
5 & \beight & Total & $1.4\pm0.3$ & $<6.4$ & \nodata & $2.3\pm0.5$ & 3.3(3) \\

\enddata
\tablecomments{All derived errors are at the 68\% confidence level.}
\tablenotetext{a} {Absorption component at the quasar or lens redshift.}
\tablenotetext{b} {Flux is estimated in the 0.4-8 keV observed band without magnification correction.}
\tablenotetext{c} {Simultaneous fits to the individual spectra of images A and B of \bzero.  The two spectra are constrained to have the same intrinsic power-law photon index and Galactic absorption column density but can have different absorption column densities at the redshift of the lens.}

\end{deluxetable}

\subsection{Flux Ratios\label{sec:fr}}
We estimated the absorption corrected flux ratio between images A and B of \bzero\ using \verb+XSPEC+ in the 0.4--8 keV band and obtained a flux ratio of B/A = $0.57\pm0.11$.
We compare this to the time delay corrected flux ratios obtained for the optical and radio bands \citep{ko00,bu00} in Table~\ref{tab:frzero}.
For \beight\ we estimated the flux ratios based on the image count rates because of the low S/N of individual images.
In addition, we estimated the count rate of combined images A and C since it is difficult to resolve the two images.
The result for \beight\ is listed in Table~\ref{tab:freight}.
The flux ratios in X-rays are consistent with the time delay corrected flux ratio in the radio band in \beight\ \citep{fa99}.

\begin{deluxetable}{cc}
\tabletypesize{\scriptsize}
\tablecolumns{2}
\tablewidth{0pt}
\tablecaption{Flux Ratios of \bzero\label{tab:frzero}}
\tablehead{
\colhead{Band} &
\colhead{B/A}
}

\startdata
X-ray ({\rm 0.4--8 keV}) & $0.57\pm0.11$ \\
Optical (I)\tablenotemark{a} & 0.67 \\
Radio ({\rm 8.5 GHz})\tablenotemark{b} & 0.825 \\
\enddata
\tablerefs{(a) \citet{bu00}; (b) \citet{ko00}}
\end{deluxetable}

\begin{deluxetable}{ccc}
\tabletypesize{\scriptsize}
\tablecolumns{3}
\tablewidth{0pt}
\tablecaption{Flux Ratios of \beight\label{tab:freight}}
\tablehead{
\colhead{Band} &
\colhead{B/(A+C)} &
\colhead{D/(A+C)} 
}

\startdata
X-ray ({\rm 0.4--8 keV}) & $0.4\pm0.1$ & $<0.12$  \\ 
Radio ({\rm 8.5 GHz})\tablenotemark{a} & 0.325 & 0.114 \\
\enddata
\tablerefs{(a) \citet{fa99}}
\end{deluxetable}

\section{Cluster Analysis}
It is important to constrain the mass and central location of any nearby groups or clusters to construct a better lens model. 
We performed source detection on the two \chandra\ observations over the full ACIS fields using the \verb+CIAO+ tool \verb+wavdetect+ \citep{fr02} provided by CXC and did not find significant extended X-ray emission in either field.
We also smoothed the image with the \verb+csmooth+ \citep{ewr00} software tool to detect any extended sources within 4\arcmin\ of the lens galaxies in the ACIS-S3 and S2 fields in both observations.  
We removed the point sources detected by \verb+wavdetect+ before the smoothing process.
However, the smoothed images do not show significant (above $3\sigma$) extended emission close to the lens galaxies in either field.
We also smoothed the soft X-ray band images between (0.4--2 keV) to increase the signal-to-noise ratio, as the X-ray emission from the cluster or group is expected to be stronger in the soft X-ray band, and did not find significant extended emission above $3\sigma$.  
We note that there are two low significance extended sources (about $2.5\sigma$) present close the lenses at R.A. = 16$^{\rm h}$ 1$^{\rm m}$ 35\seco9 and decl. = 43\degr\ 17\arcmin\ 27\arcsec\ (J2000) with a $2\sigma$ contour radius of $\sim$11\arcsec\ in the field of \bzero, and at  R.A. = 16$^{\rm h}$ 9$^{\rm m}$ 11\seco3 and decl. = 65\degr\ 32\arcmin\ 7\arcsec\ (J2000) with a $2\sigma$ contour radius of $\sim$5\arcsec\ in the field of \beight.
However, these sources have low signal-to-noise ratios and could be simply be background fluctuations.  In addition, many events in these peaks have energies below 0.4 keV where the calibration uncertainty is large.  A longer observation would be required to have any confidence in their existence, so we proceed on the assumption that they are background fluctuations.

We used the background count rates to estimate upper limits on the fluxes of any group/cluster in the two fields.  We assumed gas temperature of $\sim$1 keV when converting from count rates to flux.
This temperature consistent with the upper limits of X-ray luminosities obtained from our analysis.
The background count rate in the 0.4--2 keV band for \bzero\ is $5.8\times10^{-7}~{\rm count~s^{-1}~arcsec^{-2}}$. 
In the analysis, we used an aperture size of 100 kpc radius to limit the effects of any systematic errors in the background estimate.
This aperture size is large enough for the groups with luminosities below $10^{43}~\lumin$ \citep{op04}. 
For a 100 kpc radius at the redshift of the lens ($z=0.41$), the $3\sigma$ upper limit of the X-ray flux is $3.4\times10^{-15}~\flux$, which corresponds to an X-ray luminosity of $\sim2\times10^{42}~\lumin$
The background flux in the 0.4--2 keV band for \beight\ is $5.3\times10^{-7}~{\rm count~s^{-1}~arcsec^{-2}}$.  
For a 100 kpc radius at the redshift of the lens ($z=0.63$), the $3\sigma$ upper limit of the X-ray flux is $3.5\times10^{-15}~\flux$, which corresponds to an X-ray luminosity of $\sim6\times10^{42}~\lumin$

We used the standard $L_X$--$T_X$, $\sigma$--$T_X$, and $\sigma$--$L_X$ relations (e.g., Girardi \etal 1996; Helsdon \& Ponman 2000; Xue \& Wu 2000; Mahdavi \& Geller 2001; Kochanek \etal 2003; Osmond \& Ponman 2004) for groups and clusters to estimate upper limits for the velocity dispersion of any cluster of $\sigma\sim400$ and $\sim500$~\kms\ for \bzero\ and \beight\ respectively.
The large scatter of the relations were considered when estimating the upper limits.
Considering a singular isothermal sphere model for the group or cluster, the lens potential, $\phi$, from the group/cluster is
\begin{equation}
\phi=4\pi\left(\frac{\sigma}{c}\right)^2\frac{D_{LS}}{D_{OS}}r=br,
\end{equation}
where $D_{LS}$ and $D_{OS}$ are angular diameter distances from the lens to source and observer to source, and $r$ is the angular distance from the lens to the group/cluster.
Thus, the lens strength, b, is constrained to have upper limits of $b\ls$3\sarc0 and $\ls$3\sarc2 for \bzero\ and \beight\ respectively.

\section{Serendipitous Sources}
We also cataloged the serendipitous point sources in the two fields using the \verb+CIAO+ tool \verb+wavdetect+.
The source detection was performed on each CCD separately with a detection threshold of $10^{-6}$.  
The detection threshold was set as the inverse of the total number of pixels in the binned images.
We detected 75 and 82 serendipitous sources in the \bzero\ and \beight\ fields, respectively.
The detection threshold corresponds to a flux limit of roughly $2\times10^{-15}$~\flux\ and the numbers of sources are consistent with standard estimates for the source numbers at this flux limit \citep[e.g.,][and references there in]{co02}.

The \chandra\ field of \bzero\ is included in DR3 of the Sloan Digital Sky Survey \citep[SDSS,][]{ab04}.
We searched for optical counterparts of the X-ray sources within a 2\arcsec\ radius, which corresponds to the largest error bars for the \chandra\ positions,
and allows the inclusion of counterparts that have slight offsets between the optical and X-ray centroids, such as the ultra-luminous X-ray sources seen in nearby galaxies.
We found counterparts for 27 of the 75 X-ray sources, 13 of which are flagged as quasar candidates in the SDSS catalogs.
The properties of the serendipitous sources and the optical counterparts are listed in Table~\ref{tab:szero}.

\beight\ field is not covered by SDSS, and so we used the Palomar Digital Sky Survey \citep[DPOSS,][]{dj03} survey to identify the counterparts.  
The lens position in the \chandra\ and DPOSS images are offset by approximately 1\sarc5, so we used a slightly larger radius to identify optical counterparts.
We found 27 optical counterparts out of the 82 X-ray sources.
The properties of the serendipitous sources and the optical counterparts in \beight\ field are listed in Table~\ref{tab:seight}.

\section{Summary}
We present results from \chandra\ observations of the gravitational lenses \bzero\ and \beight. 
We resolved the two lensed images in \bzero\ and three of four images in \beight.
We did not detect the faintest image D in \beight.
The analysis of \beight\ was complicated by the small minimum image separation and the low count-rate of the lensed image D.
We could not find a stable solution for the X-ray flux ratios even if we fixed the image positions to those inferred from \hst\ observations.
A longer observation would be needed to constrain the flux ratios of the individual X-ray images of \beight.

We also performed a spectral analysis of the lensed quasar images and detected differential absorption column density of $\Delta\nh\sim3\times10^{21}\cmsq$ between the two images of \bzero.
The improvement of the fit by allowing differential absorption is significant at the 90\% confidence level indicating that the X-ray data are consistent with, but do not require, column density variations.
Comparing with the estimated differential extinction of $\Delta E(B-V)=0.1$~mag between images A and B \citep{fal99}, we obtained a dust-to-gas ratio of $2.6^{+1.9}_{-1.5}\times10^{22}~{\rm mag^{-1}~cm^{-2}}$ within a factor of five from the Galactic value.
Considering the patchiness of the interstellar medium, a factor of five difference may not be significant enough to conclude that the dust-to-gas ratio of the lens galaxy in \bzero\ is higher than that of the Milky Way.
However, this trend is consistent with the cases in B0218+357 ($z_l=0.68$) and PKS~1830--211 \citep[$z_l=0.89$,][]{fal99}, where the estimated dust-to-gas ratios in the lens galaxies were higher than the Galactic value.
It is, however, the opposite of the situation found in Q2237+0305 \citep[$z_l=0.04$,][]{da03}.
A larger sample is needed to investigate the average dust-to-gas ratios in high redshift ($z>0$) galaxies.
There is differential extinction in \beight\ as well, but it is most significant for image D which we failed to detect.
Our upper limit on the flux of D is not significant enough to estimate a limit on the dust-to-gas ratio.
The differential absorption measured between different images in gravitational lenses may serve as a convenient probe to study the gas content in high redshift galaxies since it can separate the absorbing column originated from the lens galaxy and those intrinsic to quasars.  

We did not detect significant X-ray emission from nearby galaxy groups or clusters associated with the lens galaxies.
We found upper limits on the X-ray luminosity of any cluster at the lens redshift and within 4\arcmin\ from the lenses of $\sim2\times10^{42}$ and $\sim6\times10^{42}$~\lumin\ for \bzero\ and \beight, respectively.  
Considering standard cluster relations between X-ray luminosity, temperature, and velocity dispersion and assuming a singular isothermal sphere model for the group/cluster, we obtained upper limits of the lens strength parameter, $b$, of 3\sarc0 and 3\sarc2 for \bzero\ and \beight\ respectively. 
With such a tight limit, only the poor groups to which each lens belongs could be contributing to the lens potential.

\acknowledgements
We thank Emilio Falco, Christine Jones, George Chartas, and the anonymous referee for helpful comments.
We acknowledge the financial support by \hst\ grant GO-9375 and CXC grant GO3-4154X.
\clearpage

\begin{deluxetable}{cccccccccccc}
\tabletypesize{\scriptsize}
\rotate
\tablecolumns{12}
\tablewidth{0pt}
\tablecaption{Serendipitous Sources Detected in \bzero\ Field \label{tab:szero}}
\tablehead{
\colhead{} & \colhead{} & \colhead{} & \colhead{R. A.} & \colhead{Decl.} & \colhead{} & \colhead{} & \colhead{R. A.} & \colhead{Decl.} & \colhead{} & \colhead{} & \colhead{} \\
\colhead{} & \colhead{R. A.} & \colhead{Decl.} & \colhead{error} & \colhead{error} & \colhead{Net} & \colhead{} & \colhead{optical} & \colhead{optical} & \colhead{$\rm D_{OX}$} & \colhead{r$^*$} & \colhead{} \\
\colhead{Name} & \colhead{(degree)} & \colhead{(degree)} & \colhead{(\arcsec)} & \colhead{(\arcsec)} & \colhead{Counts} & \colhead{Sig} & \colhead{(degree)} & \colhead{(degree)}  & \colhead{(\arcsec)} & \colhead{(mag)} & \colhead{SDSS flag} 
}
\startdata
CXO J160022.5+431521 &       240.09376 &       43.256002 &0.7 &0.5 &         152$\pm$          16 & 13.6 & \nodata & \nodata & \nodata & \nodata & \nodata \\
CXO J160041.2+431544 &       240.17168 &       43.262303 &1.0 &0.8 &          10$\pm$           4 &  2.8 & \nodata & \nodata & \nodata & \nodata & \nodata \\
CXO J160044.4+430842 &       240.18509 &       43.145258 &0.6 &0.2 &          26$\pm$           7 &  4.4 & \nodata & \nodata & \nodata & \nodata & \nodata \\
CXO J160045.3+431433 &       240.18909 &       43.242736 &1.1 &0.8 &          13$\pm$           4 &  3.8 &       240.18867 &       43.242565 &1.3 &      22.7490 &\verb++ \\
CXO J160047.0+431354 &       240.19615 &       43.231796 &0.5 &0.6 &           7$\pm$           3 &  2.4 & \nodata & \nodata & \nodata & \nodata & \nodata \\
CXO J160049.2+431433 &       240.20518 &       43.242667 &0.4 &1.0 &          10$\pm$           4 &  3.3 & \nodata & \nodata & \nodata & \nodata & \nodata \\
CXO J160051.2+431431 &       240.21355 &       43.242076 &1.0 &0.7 &          26$\pm$           6 &  6.3 & \nodata & \nodata & \nodata & \nodata & \nodata \\
CXO J160053.8+430932 &       240.22429 &       43.159003 &1.2 &0.7 &          28$\pm$           7 &  5.3 & \nodata & \nodata & \nodata & \nodata & \nodata \\
CXO J160054.8+431528 &       240.22854 &       43.258031 &0.9 &0.4 &          14$\pm$           4 &  4.6 &       240.22871 &       43.257572 &1.7 &      15.9220 &\verb++ \\
CXO J160055.4+431052 &       240.23113 &       43.181337 &0.8 &0.5 &           6$\pm$           3 &  2.4 & \nodata & \nodata & \nodata & \nodata & \nodata \\
CXO J160056.4+431151 &       240.23515 &       43.197578 &1.0 &0.4 &           9$\pm$           3 &  3.1 & \nodata & \nodata & \nodata & \nodata & \nodata \\
CXO J160100.8+431055 &       240.25354 &       43.182145 &0.7 &0.6 &          11$\pm$           4 &  3.5 & \nodata & \nodata & \nodata & \nodata & \nodata \\
CXO J160101.6+430108 &       240.25677 &       43.019044 &2.0 &0.8 &          12$\pm$           4 &  3.5 & \nodata & \nodata & \nodata & \nodata & \nodata \\
CXO J160104.8+431115 &       240.27041 &       43.187527 &0.9 &0.3 &           8$\pm$           3 &  2.9 & \nodata & \nodata & \nodata & \nodata & \nodata \\
CXO J160105.0+431850 &       240.27114 &       43.314013 &0.5 &0.9 &           4$\pm$           2 &  1.7 & \nodata & \nodata & \nodata & \nodata & \nodata \\
CXO J160110.8+431139 &       240.29526 &       43.194329 &0.2 &0.1 &         299$\pm$          18 & 51.1 &       240.29534 &       43.193954 &1.4 &      17.1220 &\verb+TARGET_GALAXY+ \\
CXO J160113.2+431803 &       240.30505 &       43.300870 &0.3 &0.2 &           8$\pm$           3 &  3.4 & \nodata & \nodata & \nodata & \nodata & \nodata \\
CXO J160114.4+430717 &       240.31025 &       43.121493 &0.8 &0.9 &           9$\pm$           4 &  2.7 & \nodata & \nodata & \nodata & \nodata & \nodata \\
CXO J160116.3+431740 &       240.31819 &       43.294634 &0.4 &0.2 &          11$\pm$           3 &  4.3 & \nodata & \nodata & \nodata & \nodata & \nodata \\
CXO J160116.8+431834 &       240.32011 &       43.309628 &0.2 &0.3 &          14$\pm$           4 &  5.7 & \nodata & \nodata & \nodata & \nodata & \nodata \\
CXO J160116.9+431949 &       240.32073 &       43.330454 &0.5 &0.2 &          10$\pm$           3 &  4.1 &       240.32050 &       43.330372 &0.7 &      21.8080 &\verb++ \\
CXO J160117.1+432129 &       240.32156 &       43.358059 &0.4 &0.2 &          53$\pm$           8 & 14.9 & \nodata & \nodata & \nodata & \nodata & \nodata \\
CXO J160120.8+431827 &       240.33669 &       43.307585 &0.1 &0.1 &          59$\pm$           8 & 22.9 &       240.33658 &       43.307536 &0.3 &      20.2470 &\verb+TARGET_QSO_FAINT+ \\
CXO J160121.0+430936 &       240.33777 &       43.160101 &0.3 &0.5 &           5$\pm$           2 &  2.1 & \nodata & \nodata & \nodata & \nodata & \nodata \\
CXO J160122.8+431522 &       240.34521 &       43.256365 &0.3 &0.2 &           6$\pm$           2 &  2.7 & \nodata & \nodata & \nodata & \nodata & \nodata \\
CXO J160125.6+431337 &       240.35703 &       43.227171 &0.2 &0.2 &           3$\pm$           2 &  1.5 & \nodata & \nodata & \nodata & \nodata & \nodata \\
CXO J160125.9+431832 &       240.35828 &       43.308986 &0.2 &0.1 &          24$\pm$           5 & 11.4 &       240.35824 &       43.308991 &0.1 &      21.6330 &\verb++ \\
CXO J160126.5+431704 &       240.36068 &       43.284556 &0.2 &0.1 &           8$\pm$           3 &  3.9 & \nodata & \nodata & \nodata & \nodata & \nodata \\
CXO J160126.6+431313 &       240.36091 &       43.220496 &0.3 &0.2 &          11$\pm$           3 &  5.0 & \nodata & \nodata & \nodata & \nodata & \nodata \\
CXO J160128.9+430519 &       240.37078 &       43.088780 &1.0 &0.6 &          11$\pm$           4 &  3.4 & \nodata & \nodata & \nodata & \nodata & \nodata \\
CXO J160129.2+430748 &       240.37185 &       43.130045 &0.7 &0.5 &          17$\pm$           4 &  5.9 & \nodata & \nodata & \nodata & \nodata & \nodata \\
CXO J160129.5+431940 &       240.37294 &       43.327784 &0.2 &0.2 &           7$\pm$           3 &  3.4 &       240.37304 &       43.327680 &0.5 &      14.1040 &\verb++ \\
CXO J160129.8+431533 &       240.37433 &       43.259386 &0.2 &0.1 &           9$\pm$           3 &  4.3 & \nodata & \nodata & \nodata & \nodata & \nodata \\
CXO J160130.3+431711 &       240.37637 &       43.286424 &0.0 &0.2 &           2$\pm$           1 &  1.0 & \nodata & \nodata & \nodata & \nodata & \nodata \\
CXO J160131.1+431814 &       240.37993 &       43.304018 &0.2 &0.2 &           9$\pm$           3 &  4.1 & \nodata & \nodata & \nodata & \nodata & \nodata \\
CXO J160132.7+431309 &       240.38644 &       43.219338 &0.1 &0.2 &           5$\pm$           2 &  2.4 &       240.38645 &       43.219297 &0.2 &      22.9010 &\verb++ \\
CXO J160135.3+431333 &       240.39740 &       43.226067 &0.1 &0.0 &         330$\pm$          18 &114.4 &       240.39737 &       43.226019 &0.2 &      20.4010 &\verb+TARGET_QSO_FAINT+ \\
CXO J160136.2+431353 &       240.40113 &       43.231525 &0.1 &0.3 &           4$\pm$           2 &  2.0 & \nodata & \nodata & \nodata & \nodata & \nodata \\
CXO J160136.4+432107 &       240.40199 &       43.352020 &0.1 &0.1 &          71$\pm$           9 & 28.2 &       240.40162 &       43.352124 &1.0 &      22.3800 &\verb+TARGET_QSO_REJECT+ \\
CXO J160137.2+431609 &       240.40509 &       43.269321 &0.1 &0.0 &          39$\pm$           6 & 18.5 & \nodata & \nodata & \nodata & \nodata & \nodata \\
CXO J160137.3+432236 &       240.40543 &       43.376842 &0.2 &0.2 &          23$\pm$           5 &  8.5 & \nodata & \nodata & \nodata & \nodata & \nodata \\
CXO J160138.6+432204 &       240.41088 &       43.367895 &0.4 &0.3 &           6$\pm$           3 &  2.6 & \nodata & \nodata & \nodata & \nodata & \nodata \\
CXO J160141.8+431453 &       240.42440 &       43.248288 &0.4 &0.1 &           5$\pm$           2 &  2.5 &       240.42461 &       43.248106 &0.9 &      20.9150 &\verb+TARGET_QSO_FAINT+ \\
CXO J160141.9+431453 &       240.42459 &       43.248073 &0.1 &0.1 &          23$\pm$           5 & 11.8 &       240.42458 &       43.248080 &0.0 &      20.9290 &\verb+TARGET_QSO_FAINT+ \\
CXO J160143.5+431858 &       240.43147 &       43.316195 &0.2 &0.1 &           7$\pm$           3 &  3.3 &       240.43154 &       43.316113 &0.4 &      25.7410 &\verb++ \\
CXO J160143.7+431126 &       240.43234 &       43.190683 &0.3 &0.4 &          18$\pm$           4 &  7.1 &       240.43238 &       43.190730 &0.2 &      23.3230 &\verb++ \\
CXO J160144.6+431936 &       240.43594 &       43.326831 &0.1 &0.1 &           9$\pm$           3 &  4.4 & \nodata & \nodata & \nodata & \nodata & \nodata \\
CXO J160146.4+431419 &       240.44346 &       43.238879 &0.2 &0.3 &           7$\pm$           3 &  3.5 & \nodata & \nodata & \nodata & \nodata & \nodata \\
CXO J160146.8+431334 &       240.44529 &       43.226177 &0.1 &0.1 &         223$\pm$          15 & 79.9 &       240.44533 &       43.226154 &0.1 &      21.1140 &\verb++ \\
CXO J160147.5+430907 &       240.44803 &       43.152111 &0.6 &0.9 &          12$\pm$           4 &  3.7 & \nodata & \nodata & \nodata & \nodata & \nodata \\
CXO J160147.9+432006 &       240.44998 &       43.335108 &0.1 &0.1 &          19$\pm$           4 &  8.8 & \nodata & \nodata & \nodata & \nodata & \nodata \\
CXO J160149.6+430836 &       240.45684 &       43.143382 &0.6 &0.3 &          24$\pm$           5 &  7.0 & \nodata & \nodata & \nodata & \nodata & \nodata \\
CXO J160152.6+430915 &       240.46947 &       43.154196 &0.5 &0.5 &          49$\pm$           8 & 10.9 &       240.46913 &       43.153922 &1.3 &      20.5530 &\verb+TARGET_QSO_FAINT+ \\
CXO J160153.7+431817 &       240.47391 &       43.304920 &0.0 &0.0 &         510$\pm$          23 &173.3 &       240.47392 &       43.304845 &0.3 &      18.8240 &\verb+TARGET_QSO_FAINT+ \\
CXO J160154.5+431519 &       240.47735 &       43.255280 &0.3 &0.1 &           3$\pm$           2 &  1.5 &       240.47692 &       43.255471 &1.3 &      20.6500 &\verb+TARGET_QSO_FAINT+ \\
CXO J160154.6+431147 &       240.47791 &       43.196539 &0.5 &0.5 &           6$\pm$           3 &  2.5 & \nodata & \nodata & \nodata & \nodata & \nodata \\
CXO J160154.7+431019 &       240.47811 &       43.172163 &0.2 &0.6 &           3$\pm$           2 &  1.6 &       240.47841 &       43.172037 &0.9 &      22.8610 &\verb++ \\
CXO J160156.8+431125 &       240.48683 &       43.190556 &0.2 &0.2 &         128$\pm$          12 & 32.4 &       240.48681 &       43.190463 &0.3 &      19.7180 &\verb+TARGET_QSO_FAINT+ \\
CXO J160158.0+431145 &       240.49198 &       43.195840 &0.5 &0.4 &          13$\pm$           4 &  5.3 & \nodata & \nodata & \nodata & \nodata & \nodata \\
CXO J160200.1+431228 &       240.50070 &       43.207905 &0.5 &0.3 &           7$\pm$           3 &  3.3 & \nodata & \nodata & \nodata & \nodata & \nodata \\
CXO J160203.4+431808 &       240.51433 &       43.302496 &0.2 &0.3 &           3$\pm$           2 &  1.5 & \nodata & \nodata & \nodata & \nodata & \nodata \\
CXO J160203.9+431733 &       240.51654 &       43.292745 &0.2 &0.3 &           6$\pm$           2 &  3.0 & \nodata & \nodata & \nodata & \nodata & \nodata \\
CXO J160204.2+431558 &       240.51789 &       43.266260 &0.2 &0.2 &          41$\pm$           6 & 16.8 &       240.51805 &       43.265916 &1.3 &      20.7600 &\verb++ \\
CXO J160204.3+431556 &       240.51823 &       43.265618 &0.2 &0.1 &          26$\pm$           5 & 12.2 &       240.51809 &       43.265900 &1.1 &      20.9400 &\verb++ \\
CXO J160208.0+431357 &       240.53357 &       43.232582 &0.4 &0.3 &           9$\pm$           3 &  4.1 & \nodata & \nodata & \nodata & \nodata & \nodata \\
CXO J160209.8+431208 &       240.54096 &       43.202248 &0.5 &0.5 &           9$\pm$           3 &  3.3 & \nodata & \nodata & \nodata & \nodata & \nodata \\
CXO J160210.8+430500 &       240.54508 &       43.083499 &0.7 &0.6 &          96$\pm$          15 &  7.5 &       240.54492 &       43.083590 &0.5 &      21.1700 &\verb+TARGET_QSO_FAINT+ \\
CXO J160213.0+431212 &       240.55434 &       43.203367 &0.7 &0.5 &          19$\pm$           5 &  5.4 &       240.55386 &       43.203749 &1.9 &      21.2800 &\verb+TARGET_QSO_FAINT+ \\
CXO J160217.9+430659 &       240.57464 &       43.116569 &0.5 &0.4 &          62$\pm$          12 &  6.4 & \nodata & \nodata & \nodata & \nodata & \nodata \\
CXO J160218.1+430928 &       240.57562 &       43.157988 &0.8 &0.6 &          62$\pm$           9 & 11.5 & \nodata & \nodata & \nodata & \nodata & \nodata \\
CXO J160221.4+430800 &       240.58931 &       43.133401 &0.7 &0.6 &          58$\pm$          12 &  5.9 &       240.58890 &       43.133187 &1.3 &      22.0480 &\verb++ \\
CXO J160221.5+431227 &       240.58993 &       43.207556 &0.8 &0.6 &          21$\pm$           5 &  5.9 & \nodata & \nodata & \nodata & \nodata & \nodata \\
CXO J160221.6+430319 &       240.59001 &       43.055445 &0.3 &0.2 &         903$\pm$          41 & 30.6 &       240.59003 &       43.054918 &1.9 &      17.2630 &\verb+TARGET_GALAXY TARGET_QSO_CAP+ \\
CXO J160232.1+431039 &       240.63379 &       43.177725 &0.9 &0.6 &          24$\pm$           7 &  4.0 & \nodata & \nodata & \nodata & \nodata & \nodata \\
CXO J160235.5+430813 &       240.64792 &       43.136956 &0.6 &0.5 &         162$\pm$          20 &  9.9 &       240.64794 &       43.137232 &1.0 &      20.8500 &\verb+TARGET_QSO_FAINT+ \\
\enddata
\end{deluxetable}

\clearpage

\begin{deluxetable}{ccccccccccc}
\tabletypesize{\scriptsize}
\rotate
\tablecolumns{11}
\tablewidth{0pt}
\tablecaption{Serendipitous Sources Detected in \beight\ Field \label{tab:seight}}
\tablehead{
\colhead{} & \colhead{} & \colhead{} & \colhead{R. A.} & \colhead{Decl.} & \colhead{} & \colhead{} & \colhead{R. A.} & \colhead{Decl.} & \colhead{} & \colhead{} \\
\colhead{} & \colhead{R. A.} & \colhead{Decl.} & \colhead{error} & \colhead{error} & \colhead{Net} & \colhead{} & \colhead{optical} & \colhead{optical} & \colhead{$\rm D_{OX}$} & \colhead{r} \\
\colhead{Name} & \colhead{(degree)} & \colhead{(degree)} & \colhead{(\arcsec)} & \colhead{(\arcsec)} & \colhead{Counts} & \colhead{Sig} & \colhead{(degree)} & \colhead{(degree)}  & \colhead{(\arcsec)} & \colhead{(mag)}  
}
\startdata
CXO J160713.3+652929 &       241.80581 &       65.491391 &    2.0 &    0.9 &          32$\pm$           8 &    4.4 &       241.80481 &       65.491737 &  1.9 & 21.67 \\
CXO J160719.3+652459 &       241.83052 &       65.416517 &    0.9 &    0.4 &         311$\pm$          22 &   22.3 &       241.83031 &       65.416038 &  1.8 & 19.44 \\
CXO J160727.2+653314 &       241.86370 &       65.553915 &    1.9 &    1.0 &          21$\pm$           6 &    4.2 & \nodata & \nodata & \nodata & \nodata \\
CXO J160739.8+652040 &       241.91592 &       65.344505 &    2.2 &    1.1 &          20$\pm$           6 &    3.9 & \nodata & \nodata & \nodata & \nodata \\
CXO J160740.4+652817 &       241.91870 &       65.471505 &    0.9 &    0.4 &         121$\pm$          12 &   19.8 &       241.91879 &       65.470879 &  2.3 & 21.82 \\
CXO J160741.5+653017 &       241.92300 &       65.504803 &    2.0 &    0.6 &          14$\pm$           5 &    3.6 & \nodata & \nodata & \nodata & \nodata \\
CXO J160747.1+652610 &       241.94660 &       65.436200 &    3.1 &    0.8 &          26$\pm$           7 &    5.0 &       241.94743 &       65.435570 &  2.6 & 21.08 \\
CXO J160803.3+652522 &       242.01387 &       65.422904 &    1.5 &    0.3 &           7$\pm$           3 &    2.4 & \nodata & \nodata & \nodata & \nodata \\
CXO J160804.4+652540 &       242.01872 &       65.427796 &    2.0 &    0.6 &          14$\pm$           5 &    3.8 & \nodata & \nodata & \nodata & \nodata \\
CXO J160806.5+652204 &       242.02720 &       65.367868 &    2.3 &    0.4 &          15$\pm$           5 &    4.2 & \nodata & \nodata & \nodata & \nodata \\
CXO J160807.6+652828 &       242.03167 &       65.474558 &    0.4 &    0.2 &           7$\pm$           3 &    2.7 & \nodata & \nodata & \nodata & \nodata \\
CXO J160812.0+652054 &       242.05015 &       65.348453 &    1.7 &    0.6 &          12$\pm$           4 &    3.7 &       242.05020 &       65.347809 &  2.3 &\nodata \\
CXO J160813.0+652319 &       242.05442 &       65.388855 &    1.8 &    0.6 &          16$\pm$           5 &    4.8 &       242.05438 &       65.389442 &  2.1 & 21.08 \\
CXO J160816.7+652528 &       242.06963 &       65.424496 &    1.3 &    0.7 &          13$\pm$           4 &    4.1 & \nodata & \nodata & \nodata & \nodata \\
CXO J160816.9+652453 &       242.07071 &       65.414763 &    1.7 &    0.9 &          10$\pm$           4 &    3.3 & \nodata & \nodata & \nodata & \nodata \\
CXO J160822.0+652245 &       242.09175 &       65.379180 &    1.6 &    0.7 &          20$\pm$           5 &    5.2 & \nodata & \nodata & \nodata & \nodata \\
CXO J160823.1+652328 &       242.09637 &       65.391192 &    1.2 &    0.4 &          35$\pm$           7 &    8.3 &       242.09776 &       65.390694 &  2.8 &\nodata \\
CXO J160826.7+653546 &       242.11160 &       65.596178 &    1.0 &    0.2 &          46$\pm$           8 &   10.5 &       242.11134 &       65.595993 &  0.8 & 21.84 \\
CXO J160827.3+652445 &       242.11378 &       65.412762 &    0.5 &    0.2 &           8$\pm$           3 &    3.4 & \nodata & \nodata & \nodata & \nodata \\
CXO J160831.3+652423 &       242.13047 &       65.406416 &    1.2 &    0.7 &          16$\pm$           4 &    5.2 & \nodata & \nodata & \nodata & \nodata \\
CXO J160832.8+652255 &       242.13706 &       65.381987 &    1.3 &    0.5 &          11$\pm$           4 &    3.6 & \nodata & \nodata & \nodata & \nodata \\
CXO J160835.1+653605 &       242.14658 &       65.601453 &    0.7 &    0.3 &          11$\pm$           4 &    3.9 & \nodata & \nodata & \nodata & \nodata \\
CXO J160835.8+653137 &       242.14931 &       65.527038 &    0.7 &    0.2 &           6$\pm$           2 &    2.7 &       242.15018 &       65.527252 &  1.5 & 16.37 \\
CXO J160836.5+652236 &       242.15214 &       65.376812 &    1.2 &    0.7 &          12$\pm$           4 &    3.9 & \nodata & \nodata & \nodata & \nodata \\
CXO J160840.9+653523 &       242.17070 &       65.589992 &    0.7 &    0.3 &           9$\pm$           3 &    3.3 & \nodata & \nodata & \nodata & \nodata \\
CXO J160843.8+653245 &       242.18292 &       65.546039 &    0.2 &    0.1 &          89$\pm$          10 &   35.4 &       242.18413 &       65.545868 &  1.9 & 19.86 \\
CXO J160844.1+653246 &       242.18413 &       65.546163 &    0.2 &    0.1 &          95$\pm$          10 &   39.3 &       242.18413 &       65.545868 &  1.1 & 19.86 \\
CXO J160849.1+653358 &       242.20462 &       65.566266 &    0.3 &    0.1 &          30$\pm$           6 &   13.4 & \nodata & \nodata & \nodata & \nodata \\
CXO J160850.7+653707 &       242.21163 &       65.618791 &    0.8 &    0.4 &           8$\pm$           3 &    3.4 & \nodata & \nodata & \nodata & \nodata \\
CXO J160850.8+653220 &       242.21187 &       65.539142 &    0.1 &    0.0 &         617$\pm$          25 &  207.9 &       242.21214 &       65.538887 &  1.0 & 19.78 \\
CXO J160854.0+653338 &       242.22506 &       65.560669 &    0.4 &    0.1 &          12$\pm$           3 &    5.7 & \nodata & \nodata & \nodata & \nodata \\
CXO J160855.0+652738 &       242.22933 &       65.460712 &    0.3 &    0.3 &           3$\pm$           2 &    1.5 & \nodata & \nodata & \nodata & \nodata \\
CXO J160858.9+652816 &       242.24561 &       65.471199 &    0.6 &    0.2 &           9$\pm$           3 &    4.5 & \nodata & \nodata & \nodata & \nodata \\
CXO J160859.4+653225 &       242.24789 &       65.540349 &    0.2 &    0.0 &          69$\pm$           8 &   30.0 &       242.24838 &       65.540154 &  1.0 & 21.07 \\
CXO J160900.4+653445 &       242.25186 &       65.579230 &    0.4 &    0.1 &          19$\pm$           4 &    8.7 & \nodata & \nodata & \nodata & \nodata \\
CXO J160901.6+653744 &       242.25685 &       65.629054 &    1.3 &    0.3 &          21$\pm$           5 &    6.5 & \nodata & \nodata & \nodata & \nodata \\
CXO J160902.1+653216 &       242.25881 &       65.537796 &    0.7 &    0.1 &           4$\pm$           2 &    1.9 & \nodata & \nodata & \nodata & \nodata \\
CXO J160903.9+653040 &       242.26625 &       65.511286 &    0.4 &    0.2 &           8$\pm$           3 &    3.6 & \nodata & \nodata & \nodata & \nodata \\
CXO J160904.8+652737 &       242.27018 &       65.460319 &    0.2 &    0.1 &          25$\pm$           5 &   11.8 & \nodata & \nodata & \nodata & \nodata \\
CXO J160905.2+653655 &       242.27176 &       65.615366 &    1.0 &    0.2 &          13$\pm$           4 &    4.6 & \nodata & \nodata & \nodata & \nodata \\
CXO J160905.5+652400 &       242.27325 &       65.400242 &    1.6 &    0.4 &           6$\pm$           3 &    2.3 & \nodata & \nodata & \nodata & \nodata \\
CXO J160906.1+652700 &       242.27560 &       65.450274 &    0.7 &    0.4 &          14$\pm$           4 &    5.7 & \nodata & \nodata & \nodata & \nodata \\
CXO J160907.1+653126 &       242.27985 &       65.524103 &    0.3 &    0.0 &           6$\pm$           2 &    2.9 &       242.28047 &       65.523933 &  1.1 & 20.14 \\
CXO J160907.9+652626 &       242.28297 &       65.440657 &    0.5 &    0.3 &          16$\pm$           4 &    7.0 &       242.28284 &       65.440109 &  2.0 & 21.62 \\
CXO J160909.1+652559 &       242.28817 &       65.433131 &    0.8 &    0.3 &           6$\pm$           2 &    2.7 & \nodata & \nodata & \nodata & \nodata \\
CXO J160909.6+653252 &       242.29026 &       65.547984 &    0.3 &    0.1 &          17$\pm$           4 &    8.1 & \nodata & \nodata & \nodata & \nodata \\
CXO J160911.0+653008 &       242.29611 &       65.502295 &    0.4 &    0.1 &           5$\pm$           2 &    2.4 & \nodata & \nodata & \nodata & \nodata \\
CXO J160913.0+653432 &       242.30456 &       65.575761 &    0.2 &    0.1 &          92$\pm$          10 &   38.9 &       242.30478 &       65.575493 &  1.0 &\nodata \\
CXO J160913.1+652457 &       242.30483 &       65.415846 &    1.3 &    0.3 &           5$\pm$           2 &    2.3 & \nodata & \nodata & \nodata & \nodata \\
CXO J160913.1+653020 &       242.30487 &       65.505795 &    0.3 &    0.2 &           6$\pm$           2 &    2.8 & \nodata & \nodata & \nodata & \nodata \\
CXO J160913.4+652827 &       242.30590 &       65.474394 &    0.1 &    0.1 &          82$\pm$           9 &   36.7 &       242.30664 &       65.474052 &  1.7 & 20.42 \\
CXO J160914.1+653700 &       242.30883 &       65.616860 &    0.7 &    0.3 &          18$\pm$           5 &    6.5 & \nodata & \nodata & \nodata & \nodata \\
CXO J160914.4+653206 &       242.31012 &       65.535131 &    0.2 &    0.1 &          39$\pm$           6 &   17.8 &       242.31021 &       65.534958 &  0.6 &\nodata \\
CXO J160916.8+652852 &       242.32012 &       65.481128 &    0.2 &    0.1 &           8$\pm$           3 &    4.1 & \nodata & \nodata & \nodata & \nodata \\
CXO J160917.4+653021 &       242.32254 &       65.506092 &    0.2 &    0.1 &          29$\pm$           5 &   13.9 & \nodata & \nodata & \nodata & \nodata \\
CXO J160917.4+652925 &       242.32268 &       65.490353 &    0.0 &    0.0 &           2$\pm$           1 &    1.0 & \nodata & \nodata & \nodata & \nodata \\
CXO J160917.7+653104 &       242.32390 &       65.517999 &    0.3 &    0.1 &           7$\pm$           3 &    3.4 & \nodata & \nodata & \nodata & \nodata \\
CXO J160919.6+652915 &       242.33176 &       65.487717 &    0.5 &    0.1 &           7$\pm$           3 &    3.5 &       242.33199 &       65.487556 &  0.7 & 18.87 \\
CXO J160919.9+653552 &       242.33310 &       65.597953 &    0.4 &    0.1 &          29$\pm$           6 &   11.2 & \nodata & \nodata & \nodata & \nodata \\
CXO J160921.4+652243 &       242.33930 &       65.378676 &    1.4 &    0.6 &          11$\pm$           4 &    3.4 &       242.34024 &       65.378616 &  1.4 & 17.14 \\
CXO J160925.2+653421 &       242.35506 &       65.572583 &    0.1 &    0.1 &          36$\pm$           6 &   17.0 & \nodata & \nodata & \nodata & \nodata \\
CXO J160926.6+653844 &       242.36086 &       65.645732 &    0.6 &    0.2 &         180$\pm$          14 &   32.6 &       242.36060 &       65.645737 &  0.4 & 20.59 \\
CXO J160927.2+653247 &       242.36361 &       65.546655 &    0.3 &    0.1 &           5$\pm$           2 &    2.4 &       242.36388 &       65.546600 &  0.4 & 21.39 \\
CXO J160930.8+652810 &       242.37839 &       65.469570 &    0.3 &    0.0 &           4$\pm$           2 &    2.1 & \nodata & \nodata & \nodata & \nodata \\
CXO J160936.4+652906 &       242.40200 &       65.485094 &    0.3 &    0.1 &           6$\pm$           2 &    3.0 & \nodata & \nodata & \nodata & \nodata \\
CXO J160938.3+653447 &       242.40970 &       65.579868 &    0.2 &    0.1 &          79$\pm$           9 &   30.2 &       242.40987 &       65.579842 &  0.3 & 20.33 \\
CXO J160939.7+653235 &       242.41543 &       65.543118 &    0.1 &    0.1 &         105$\pm$          10 &   44.3 &       242.41559 &       65.542908 &  0.8 & 20.73 \\
CXO J160940.1+653313 &       242.41720 &       65.553738 &    0.3 &    0.0 &           4$\pm$           2 &    1.9 & \nodata & \nodata & \nodata & \nodata \\
CXO J160941.5+652813 &       242.42317 &       65.470465 &    0.4 &    0.3 &           7$\pm$           3 &    3.3 & \nodata & \nodata & \nodata & \nodata \\
CXO J160943.6+653317 &       242.43203 &       65.554966 &    0.4 &    0.2 &          16$\pm$           4 &    7.6 &       242.43243 &       65.554749 &  1.0 &\nodata \\
CXO J160943.8+652700 &       242.43287 &       65.450271 &    0.8 &    0.0 &           2$\pm$           1 &    1.0 & \nodata & \nodata & \nodata & \nodata \\
CXO J160948.2+653330 &       242.45110 &       65.558338 &    0.6 &    0.1 &           6$\pm$           2 &    2.7 & \nodata & \nodata & \nodata & \nodata \\
CXO J160950.9+652922 &       242.46227 &       65.489607 &    0.3 &    0.0 &           4$\pm$           2 &    2.0 & \nodata & \nodata & \nodata & \nodata \\
CXO J160953.0+653155 &       242.47116 &       65.531966 &    0.4 &    0.2 &           4$\pm$           2 &    1.8 & \nodata & \nodata & \nodata & \nodata \\
CXO J160956.5+652824 &       242.48574 &       65.473340 &    0.9 &    0.1 &           7$\pm$           3 &    3.1 & \nodata & \nodata & \nodata & \nodata \\
CXO J160957.6+652129 &       242.49040 &       65.358242 &    0.6 &    0.2 &          18$\pm$           7 &    2.8 & \nodata & \nodata & \nodata & \nodata \\
CXO J161001.4+652845 &       242.50600 &       65.479256 &    0.4 &    0.2 &          71$\pm$           9 &   24.3 &       242.50522 &       65.479378 &  1.3 & 18.89 \\
CXO J161004.8+652454 &       242.52003 &       65.415231 &    0.6 &    0.9 &           3$\pm$           2 &    1.5 & \nodata & \nodata & \nodata & \nodata \\
CXO J161010.8+652719 &       242.54515 &       65.455479 &    0.4 &    0.0 &           5$\pm$           2 &    2.1 & \nodata & \nodata & \nodata & \nodata \\
CXO J161016.6+652354 &       242.56936 &       65.398384 &    1.6 &    0.8 &           8$\pm$           3 &    2.7 & \nodata & \nodata & \nodata & \nodata \\
CXO J161029.9+652405 &       242.62477 &       65.401536 &    0.9 &    0.5 &         142$\pm$          14 &   17.7 &       242.62561 &       65.401230 &  1.7 & 20.03 \\
CXO J161052.3+651833 &       242.71824 &       65.309300 &    1.2 &    0.8 &          83$\pm$          14 &    7.5 & \nodata & \nodata & \nodata & \nodata \\
\enddata
\end{deluxetable}

\end{document}